\documentclass[12pt]{article}
\title{Theta and Riemann xi function representations from harmonic oscillator
eigensolutions}
\author{Mark W. Coffey\\
Department of Physics\\
Colorado School of Mines\\
Golden, CO  80401\\
(Received $\mbox{~~~~~~~~~~~~~~~~~~~~~~~~~~~~~~~2006}$)}
\date{October 4, 2006}
\begin{document}
\maketitle
\baselineskip=25 pt
\begin{abstract}

From eigensolutions of the harmonic oscillator or Kepler-Coulomb Hamiltonian 
we extend the functional equation for the Riemann zeta function and develop
integral representations for the Riemann xi function that is the completed
classical zeta function.  A key result provides a basis for generalizing the
important Riemann-Siegel integral formula.  

\end{abstract}

\vfill
\baselineskip=15pt
\centerline{\bf Key words and phrases}
\medskip

\noindent
harmonic oscillator, Mellin transformation, Hermite polynomial, hypergeometric series, functional equation, theta function, Riemann xi function 

\medskip
\vspace{.25cm}
\centerline{\bf PACS classification numbers}
02.30.Gp, 02.30.Uu, 02.30.-f

\bigskip
\centerline{\bf AMS classification numbers}
33C05, 42C05, 44A15, 44A20

\bigskip
\centerline{\bf Author contact information}
fax 303-273-3919, e-mail mcoffey@mines.edu

\baselineskip=25pt
\pagebreak
\medskip
\centerline{\bf Introduction}
\medskip

The quantum harmonic oscillator and Kepler-Coulomb problems are of enduring
interest to mathematical physics and are in a sense the same problem, for there
are many ways to transform one to the other.  For example, the 4-dimensional (8-dimensional) harmonic oscillator may be transformed to a 3-dimensional (5-dimensional) Coulomb problem \cite{ks,lambert}.  Many more mappings are realizable, especially for the corresponding radial problems \cite{ks}. 
Therefore, the eigensolutions are very closely related for these two problems
with central potentials.

We recall that the eigensolution of the fundamental quantum mechanical 
problems of the harmonic oscillator and hydrogenic atoms contain Hermite or
associated Laguerre polynomials, depending upon the coordinate system used and
the spatial dimension (e.g., \cite{nieto,coffeyjpa}).  Two quantum numbers appear
for indexing the energy levels and the angular momentum.  These wavefunctions
have a wide variety of applicability, including to image processing and the
combinatorics of zero-dimensional quantum field theory \cite{coffeyjpa}.  
Very recently additional analytic properties of these "quantum shapelets" have
been expounded \cite{coffeyjpa}.  The one-dimensional Coulomb problem has recently
reappeared as a model in quantum computing with electrons on liquid helium
films \cite{platzman,nieto2}.  
In addition, given the self-reciprocal Fourier transform property of Hermite
polynomials, there are several applications in Fourier optics 
\cite{coffey1994,horikis}.
Hermite and Laguerre polynomials are also important in random matrix theory,
especially for determinantal processes \cite{vf,mehta}.  There, the joint
probability density function of the eigenvalues of a matrix from one of the
Gaussian invariant ensembles is proportional to a Vandermonde determinant and
the orthogonal polynomials form the kernel giving the $n$-point correlation
function.

The harmonic oscillator Hamiltonian may be given a group theoretic interpretation via the Weil representation of SL$_2$(R) \cite{bumpchoi,kurlberg} and such
theory together with that of quantum mechanical commutation relations figures
prominently in constructing theta functions \cite{weyl,cartier}.  From the
Mellin transform of suitable theta functions, one may represent completed
zeta functions and one of our main results demonstrates this.  Additionally,
we present representations of the Riemann xi function that offer the
possibility to generalize the very important Riemann-Siegel formula in the
theory of the Riemann zeta function \cite{edwards,titch}.

The work of Bump et al. \cite{bumpng,bumpchoi,kurlberg} on the Mellin transforms
of Hermite and associated Laguerre functions has created significant interest  since the zeros of these functions lie only on the critical line Re $s=1/2$. 
A complementary point of view is possible within the theory of
special functions and in particular, when the Mellin transforms
are written in terms of the Gauss hypergeometric function, well known 
transformation formulae yield functional equations, reciprocity laws, and other
properties \cite{coffeyjpa2006}.  As an example, we have very recently shown
that the Mellin transform of the Laguerre function ${\cal{L}}_n^\alpha(x)=
x^{\alpha/2}e^{-x/2}L_n^{\alpha}(x)$, where $\alpha >-1$ and $L_n^\alpha$
is the associated Laguerre polynomial \cite{andrews,grad,lebedev}, is given
by $M_n^{\alpha}(s)=2^{s+\alpha/2}\Gamma(s+\alpha/2)P_n^{\alpha}(s)$ where
$\Gamma$ is the Gamma function and $P_n^{\alpha}(s)={{(1+\alpha)_n} \over {n!}}
~_2F_1(-n,s+\alpha/2;\alpha+1;2)$, with $_pF_q$ the generalized hypergeometric
function and $(a)_n=\Gamma(a+n)/\Gamma(a)$ the Pochhammer symbol.  In fact,
the polynomials $P_n^{\alpha}$ are closely related to the symmetric 
Meixner-Pollaczek polynomials $P_n^{(\lambda)}(x,\pi/2)$ \cite{atak,koe}, and
the latter are very useful in random matrix theory \cite{vf}.   

Keating \cite{keating} showed how to generalize Riemann's second proof of
the functional equation of $\zeta(s)$ by using Mellin and Fourier transforms
of Hermite polynomials.  In the next section we illustrate the hypergeometric
function point of view of those results.  The succeeding section contains our
key result.  We extend earlier work \cite{bumpng} such that the representations
given for the xi function could provide a method to generalize the 
Riemann-Siegel integral formula of complex analysis.  Again we emphasize that
these representations are built from the eigensolutions of self-adjoint
operators, these operators being the distinguished harmonic oscillator or
Coulomb Hamiltonians.

\medskip
\centerline{\bf A family of zeta functions}
\medskip

In Ref. \cite{keating} Keating generalized the Jacobi inversion formula, 
applying Poisson summation to Hermite functions ${\cal{H}}_n(x)=x^\ell H_n(ax)
e^{-cx^2}$, where $H_n$ is the Hermite polynomial, $a \neq 0$, $c=a^2/2$, and
$\ell =0, 1, 2, \dots$.  Let the $\theta$ function $\omega_{n,\ell}(t)$ be given by
$$\omega_{n,\ell}(t)=\sum_{m=1}^\infty m^\ell H_n(\sqrt{2\pi t}m)e^{-\pi m^2 t}.
\eqno(1)$$
Then we have 
{\newline \bf Theorem.} \cite{keating} For $n=2q$, $n+\ell=2p$, $p, q, \ell=
0,1,2,\ldots, \mbox{Re}~ s >1+\ell$, 
$$\int_0^\infty \omega_{n,\ell}(t)t^{s/2-1}dt = \pi^{-s/2}\Gamma(s/2)\zeta_{n,\ell}
(s), \eqno(2)$$
where 
$$\zeta_{2q,\ell}(s)=(2q)!\zeta(s-\ell)\left[H_q(s)+{{(-1)^q} \over {q!}}\right],
\eqno(3)$$
and
$$H_q(s) \equiv \sum_{k=0}^{q-1} {{(-1)^k 2^{3(q-k)}} \over {k!(2q-2k)!}}
\left({s \over 2}\right )\left({s \over 2}+1\right )\cdots\left({s \over 2}+
q-k-1 \right).  \eqno(4)$$
This theorem states that a Mellin transform of $\omega_{n,\ell}$ gives rise to
a family of zeta functions.  In this section we first derive the terminating 
hypergeometric series for the polynomial factor of Eq. (2).  We have
{\newline \bf Proposition 1}.  For $q$ a nonnegative integer
$$P_q(s) \equiv H_q(s)+{{(-1)^q} \over {q!}}= {{(-1)^q} \over {q!}} ~_2F_1\left(-q,{s \over 2};{1 \over 2};2\right).  \eqno(5)$$
Once this form of $P_q(s)$ is obtained, one may determine the functional
equation $P_q(s)=(-1)^q P_q(1-s)$ and that its zeros are all simple and lie
only on Re $s = 1/2$ \cite{bumpng,coffeyjpa2006}.

Proof of Proposition 1.  We begin by writing
$$H_q(s) \equiv \sum_{k=0}^{q-1} {{(-1)^k 2^{3(q-k)}} \over {k!(2q-2k)!}}
\left({s \over 2}\right )_{q-k}, \eqno(6)$$
and using $1/(x-n)!=(-1)^n (-x)_n/\Gamma(x+1)$, $(x)_{n-k}=(-1)^k(x)_n/
(1-x-n)_k$, and the duplication formula $(x)_{2n}=2^{2n}(x/2)_n ((x+1)/2)_n$.
These steps give
$$H_q(s)={{(s/2)_q2^{3q}} \over {(2q)!}}\sum_{k=0}^{q-1} {{(-q)_k(1/2-q)_k}
\over {(1-s/2-q)_k}} {1 \over {k!}}{1 \over 2^k}, \eqno(7)$$
from which we may read off
$$H_q(s)={{(s/2)_q2^{3q}} \over {(2q)!}}\left[~_2F_1(-q,1/2-q;1-s/2-q;1/2)
-{{(-q)_q(1/2-q)_q} \over {(1-s/2-q)_q}} {1 \over {q!}}{1 \over 2^q} \right ].
\eqno(8)$$
By using $(-q)_q=(-1)^q q!$ we then have by re-arranging the last term of 
this equation
$$H_q(s)={{(s/2)_q2^{3q}} \over {(2q)!}}\left[~_2F_1(-q,1/2-q;1-s/2-q;1/2)
\right.$$
$$\left. -\left(-{1 \over 2}\right)^q {{\sin(\pi s/2)} \over \sqrt{\pi}}{{\Gamma(s/2)} \over {\Gamma(1/2-q)}} \Gamma(1-s/2-q)\right ].  \eqno(9)$$
The $_2F_1$ function of Eq. (9) can be transformed to another hypergeometric
function at argument $-1/2$, $-1$, or $2$.  In particular, if we apply
(\cite{grad}, 9.132.2) we have
$$~_2F_1(-q,1/2-q;1-s/2-q;1/2)={{\Gamma(q+1/2)} \over {\pi^{3/2}2^q}}\Gamma(
1-s/2-q)\sin(\pi s/2)\Gamma(s/2)~_2F_1(-q,s/2;1/2;2) \eqno(10a)$$ 
$$={{(-1)^q \Gamma(q+1/2)} \over {\sqrt{\pi} 2^q\Gamma(s/2+q)}}\Gamma(s/2)
~_2F_1(-q,s/2;1/2;2). \eqno(10b)$$
In obtaining Eq. (10a) we used $\Gamma(1/2-q)\Gamma(1/2+q)=\pi/\cos(\pi q)
=(-1)^q \pi$ for $q$ a nonnegative integer.  By then inserting Eq. (10b) 
into Eq. (9) and re-arranging we have
$$H_q(s)={{2^{2q}} \over {(2q)!}} {{\Gamma(q+1/2)} \over \sqrt{\pi}} (-1)^q
\left [~_2F_1(-q,s/2;1/2;2)-1\right]. \eqno(11)$$
We then use $\Gamma(q+1/2)=\sqrt{\pi}(2q-1)!!/2^q$ and $(2q-1)!!=(2q)!/2^q q!$
to arrive at
$$H_q(s)={{(-1)^q} \over {q!}}\left[~_2F_1(-q,s/2;1/2;2)-1\right],  \eqno(12)$$
giving Eq. (5).

{\bf Corollary 1}.  By applying the transformation formula \cite{grad}
$_2F_1(\alpha,\beta;\gamma;z)=(1-z)^{-\alpha} ~_2F_1\left (\alpha,\gamma-\beta;\gamma;{z \over {z-1}}\right )$ we immediately find the
functional equation $P_q(s)=(-1)^q P_q(1-s)$.

{\bf Corollary 2}.  From the functional equation we have $P_q^{(j)}(s)
=(-1)^{q+j}P_q^{(j)}(1-s)$ so that $P_q^{(j)}(1/2)=0$ when $q+j$ is an odd
integer and in particular $P_q(1/2)=0$ when $q$ is odd.  The latter fact
may also be found from the $\alpha=-1/2$ case of the answer to the following question.
For what values of $n$ is $_2F_1(-n,(\alpha+1)/2;\alpha+1;2)=0$?  The
hypergeometric function here may be written as an $n$th divided difference
summation and as a result $_2F_1(-n,(\alpha+1)/2;\alpha+1;2)=\Gamma(1/2)
\Gamma[-(n+\alpha)/2]/\Gamma(-\alpha/2)\Gamma[(1-n)/2]$.  The simple poles
of the last $\Gamma$ factor dictate that the $_2F_1$ function vanishes
when $n$ is an odd integer, as expected.

The following subsumes Theorem 3.4 of Keating \cite{keating} for special
values of the function $\zeta_{2q,0}(s)$.  
{\newline \bf Proposition 2}.  Put $c_q \equiv (-1)^q(2q)!/q!=H_{2q}(0)$
and let $B_j$ denote the Bernoulli numbers.  For $n, ~q=0,1,2,\ldots$ and $m=1,2,\ldots$ we have 
$$\mbox{(i)}~~~~~~~~~~~~~~~~~~~~~~~~~
\zeta_{2q,0}(s)=c_q ~_2F_1(-q,-s/2;1/2;2)\zeta(s), ~~~~~~~~~~~~~~~~~~~~~~~~~
~~~~~~$$
$$\mbox{(ii)} ~~~~~~~~~~~~~~~
\zeta_{2q,0}(2m)=c_q ~_2F_1(-q,-m;1/2;2)(2\pi)^{2m}{{(-1)^{m+1} B_{2m}}
\over {2(2m)!}}, ~~~~~~~~~~~~~~~~~~~$$
$$\mbox{(iii)} ~~~~~~~~~~~~~~~~~~~~~~ 
\zeta_{2q,0}(-n)=c_q ~_2F_1(-q,n/2;1/2;2){{(-1)^n B_{n+1}} \over {n+1}},
~~~~~~~~~~~~~~~~~~~~~~~~$$
and
$$\mbox{(iv)} ~~~~~~~~~~~~~
\zeta_{2q,0}'(s)=c_q\left[_2F_1(-q,s/2;1/2;2)\zeta'(s)+\zeta(s) {d \over {ds}} ~_2F_1(-q,s/2;1/2;2)\right], ~~~~~~~~~~~$$
where 
$${d \over {ds}}~_2F_1(-q,s/2;1/2;2)={1 \over 2}\sum_{j=1}^q{{(-q)_j} \over {(1/2)_j}}
(s/2)_j\left[\psi(s/2+j)-\psi(s/2)\right ] {2^j \over {j!}}$$
$$={1 \over 2}\sum_{j=1}^q{{(-q)_j} \over {(1/2)_j}}(s/2)_j\sum_{k=0}^{j-1} {1 \over
{(s/2+k)}} {2^j \over {j!}}, \eqno(13)$$
and $\psi = \Gamma'/\Gamma$ is the digamma function \cite{andrews,nbs,grad}.
In particular, we have the values $\zeta(0)=-1/2$, $\zeta'(0)=-{1 \over 2}
\ln 2\pi$ and (v)
$$\left.{d \over {ds}}~_2F_1(-q,s/2;1/2;2)\right|_{s=0}={1 \over 2} \sum_{j=1}^q
{{(-q)_j} \over {(1/2)_j}} {2^j \over j}=-2q ~_3F_2(1-q,1,1;3/2,2;2). 
\eqno(14)$$

Proof of Proposition 2.  Parts (i), (ii), and (iii) are obvious from Eqs. (3)
and (5) and the value of the Riemann zeta function at the negative integers or
at the positive even integers.  Part (iii) follows by writing the series form
of the function $_2F_1$ and using the derivative of the Pochhammer symbol
$(d/dz)(a)_n=(a)_n[\psi(a+n)-\psi(a)]$.  The second line of Eq. (13) follows
by applying the functional equation of the digamma function \cite{grad}.

Part (v) can be obtained as the limit $s \to 0$ of the result of (iv).  We
provide another proof using a representation of  the Pochhammer symbol
in terms of Stirling numbers of the first kind $s(j,k)$ \cite{nbs}.
We have 
$$(z)_n=\prod_{k=1}^n (z+k-1)= \sum_{k=0}^n (-1)^{n-k} s(n,k)z^k.  \eqno(15)$$
Therefore we have $(d/dz)(z)_n|_{z=0}=(n-1)!$, where we used $s(n,1)=(-1)^{n-1}
(n-1)!$.  Using this fact in the series form of $_2F_1$ gives the first equality
in Eq. (14).  For the second equality in Eq. (14) we first shift the summation
index in the previous one:
$$\left.{d \over {ds}}~_2F_1(-q,s/2;1/2;2)\right|_{s=0}=\sum_{j=0}^{q-1}
{{(-q)_{j+1}} \over {(1/2)_{j+1}}} {2^j \over {(j+1)}}.  \eqno(16)$$
We then use $1/(j+1)=(1)_j/(2)_j$, apply the property $(a)_{j+1}=a(a+1)_j$,
and the rest of Eq. (14) follows.

{\bf Remarks}.  Part (iii) of the Proposition covers the trivial zeros of the
zeta function when $n=2m$ is an even integer.  In regard to part (iv), 
$\zeta'(s)$ is often easily found in terms of $\zeta(s)$ itself from the
functional equation.  For instance, we have $\zeta'(-2n) =(-1)^n (2n)!
\zeta(2n+1)/2(2\pi)^{2n}$ for $n=1,2,\ldots$.  By the method of part (iv) 
higher derivatives of $\zeta_{2q,0}(s)$ may be computed.  

\medskip
\centerline{\bf Representation of the Riemann xi function}
\medskip

Similarly to the $\ell=0$ case of Eq. (1), we put $\theta_j(x) \equiv
\sum_{n=-\infty}^\infty f_{2j}(n\sqrt{x})$ where $f_n(x)=(8\pi)^{-n/2} H_n(
\sqrt{2\pi}x)e^{-\pi x^2}$.  We then put
$$\psi_j(x)={1 \over 2}[\theta_j(x)-f_{2j}(0)]=\sum_{n=1}^\infty f_{2j}
(n\sqrt{x}),  \eqno(17)$$ 
where explicitly \cite{grad} $f_{2j}(0)=(-1)^j (4\pi)^{-j}(2j-1)!!$.
The functional equation of $\theta_j$ carries over to $\psi_j$ so that
the following fact is essentially given in Ref. \cite{bumpng}.
{\newline \bf Lemma 1}.  The function $\psi_j$ satisfies
$$\psi_j(x)={{(-1)^j} \over \sqrt{x}} \psi_j\left({1 \over x}\right)+
{1 \over 2}\left[{{(-1)^j} \over \sqrt{x}}-1\right]f_{2j}(0).  \eqno(18)$$

We let as usual $\xi$ be the Riemann xi function, defined for all complex
$s$ as $\xi(s)=s(s-1)\pi^{-s/2}\Gamma(s/2)\zeta(s)/2$ and satisfying 
$\xi(s)=\xi(1-s)$.  Very closely related to Eq. (5), we put $p_n(s)=(8\pi)^{-n}(-1)^n(2n)! ~_2F_1(-n,s/2;1/2;2)/n!$.
We then have Propositions 3 and 4.

{\bf Proposition 3}.  For $b>0$ there holds
$$p_j(s)  {{2 \xi(s)} \over {s(s-1)}} =(-1)^j\int_{1/b}^\infty x^{-(s+1)/2}
\psi_j(x)dx+\int_b^\infty \psi_j(x) x^{s/2-1}dx$$
$$-f_{2j}(0)\left[{b^{s/2} \over
s} + {{(-1)^j} \over {1-s}}b^{(s-1)/2}\right].  \eqno(19)$$
{\newline \bf Proposition 4}.  For complex $b$ with $|\mbox{arg}~ b| < \pi/2$
we have (i)
$$p_j(s)  {{2 \xi(s)} \over {s(s-1)}} = F_b(s) + (-1)^j F_{b^{-1}}(1-s)
\eqno(20)$$
and (ii) when additionally $|b|=1$ we have
$$p_j(s)  {{2 \xi(s)} \over {s(s-1)}} = F_b(s) + (-1)^j \overline{F_b(1-\overline{s})},  \eqno(21)$$
where we define
$$F_b(s) = \int_b^\infty \psi_j(x)x^{s/2-1}dx - f_{2j}(0) {b^{s/2} \over s}.
\eqno(22)$$

When $j=0$, $f_0(0)=1$ and Proposition 3 reduces to the original equation of
Riemann \cite{riemann}, while for the special case $b=1$ a result of Bump and
Ng \cite{bumpng} and Keating \cite{keating} (Theorem 3.2) is recovered.  

For the proof of Proposition 3 we alternatively evaluate the integral
$$\int_0^\infty \psi_j(x) x^{s/2-1}dx=p_j(s)\pi^{-s/2}\Gamma(s/2)\zeta(s)  
\eqno(23)$$
as $\int_0^\infty \psi_j(x) x^{s/2-1}dx=\int_0^b \psi_j(x) x^{s/2-1}dx +
\int_b^\infty \psi_j(x) x^{s/2-1}dx$.  We then put $y=1/x$ in the integral
on $[0,b]$, apply Lemma 1, and the result Eq. (19) follows.  

Proof of Proposition 4.  (i)  Equation (19) holds not just for positive $b$
but for all values $b$ in the wedge $|\mbox{arg} ~b| < \pi/2$ where the
theta function $\psi_j(x)$ is well defined.  (ii) The complex conjugate of the
function $F_b(s)$ of Eq. (18) is $F_{\overline{b}}(\overline{s})$ so that Eq. 
(21) holds whenever $\overline{b}=b^{-1}$.  That is, part (ii) holds 
whenever $b$ lies on the unit circle between $-i$ and $i$. 

\medskip
\centerline{\bf Final remarks}
\medskip 

(i) The speciality of the points $x=\pm i$ for the function $\psi_j(x)$ can
be seen by writing
$$\psi_j(x)=(8\pi)^{-j}\sum_{n=1}^\infty H_{2j}(\sqrt{2\pi x}n)e^{-\pi n^2 x}
=(8\pi)^{-j}\sum_{n=1}^\infty H_{2j}(\sqrt{2\pi x}n)e^{-\pi n^2 i}e^{-\pi n^2 (x-i)}$$
$$=(8\pi)^{-j}\sum_{n=1}^\infty H_{2j}(\sqrt{2\pi x}n)(-1)^ne^{-\pi n^2 (x-i)}$$
$$=-(8\pi)^{-j}\sum_{\stackrel{n=1}{n odd}}^\infty H_{2j}(\sqrt{2\pi x}n)e^{-\pi n^2 (x-i)}
+(8\pi)^{-j}\sum_{k=1}^\infty H_{2j}(\sqrt{2\pi x}2k)e^{-4\pi k^2 (x-i)}.
\eqno(24)$$
(ii) The properties of a theta function have also been useful in an inverse
scattering approach to the Riemann hypothesis (e.g. \cite{cofpla}).
(iii) Proposition 4 offers a prospect for generalizing the very important
Riemann-Siegel integral formula \cite{edwards} that takes the form
$${{2 \xi(s)} \over {s(s-1)}} = F(s) + \overline{F(1-\overline{s})}, \eqno(25)$$
where $F(s)=\pi^{-s/2}\Gamma(s/2)I(s)$ and $I(s)$ is a certain contour integral
taken over a straight line of slope $-1$ crossing the real axis between $0$ and
$1$ and directed from upper left to lower right.



\medskip

\centerline{\bf Acknowledgement}
This work was partially supported by Air Force contract number FA8750-06-1-0001.

\pagebreak

\end{document}